# PHOTONIC LENSES WITH WHISPERING GALLERY WAVES AT JANUS PARTICLES

Igor V. Minin[1], Oleg V, Minin[1], Yinghui Cao[2], Bing Yan[3], Zengbo Wang[3], Boris Luk'yanchuk[4]

[1]Tomsk Polytechnic University, 36 Lenin Avenue, Tomsk 634050, Russia
[2]College of Computer Science and Technology, Jilin University, 2699 Qianjin Street, Changchun 130012, China
[3]School of Electronic Engineering, Bangor University, Dean Street, Bangor, Gwynedd, LL57 1UT, UK.
[4]Faculty of Physics, Lomonosov Moscow State University, Moscow, 119991, Russia.

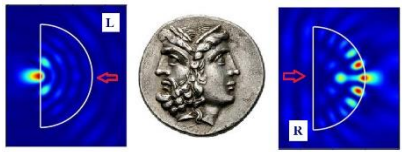

*I find among doctors that those places that should be cauterized are best considered cauterized with a crystal ball, holding it against the rays of the sun.*

Pliny the Elder, *Natural History*, X.28 (AD 77)

**Annotation:** We show that electric field on the plane surface of truncated sphere or cylinders (so called Janus particles) have sharp resonances versus the depth of removed segment of a sphere or cylinder. These resonances are related to the excited whispering gallery waves caused by truncation. It is a new mechanism of the field localization. Optimization of this effect for cylinders permits to reach a super resolution in the line thickness what can be used for contacting optical lithography.

Development of optical lithography with nanoscale resolution has been a long-standing goal for the nanotechnology community [1]. Among the many suggested methods there were a lithography with transparent dielectric particles which were used as a microscopic lenses [2-5]. The basic physical features of this technique can be well understood from the Mie theory [6] and also from the more complicated problem "particle on surface" [7, 8]. According to these calculations a transparent dielectric sphere with a diameter of more than three wavelengths of radiation incident on it can function either as a focusing lens (photon jet mode) or as a resonator concentrating energy in the whispering gallery waves in the wall region [9]. A similar behavior is shown by a transparent cylinder. The transition from the photon jet mode to the resonator mode with whispering gallery waves occurs when the size of the sphere or the radiation wavelength changes. Both phenomena are perfectly described in the framework of Mie theory, see e.g. [10]. New phenomena arise in particles in which a segment of a sphere or cylinder is removed (Janus particles [11,12]). It is a typical design of solid immersion lens [13,14]. It is known that parameters of a photon jet from a hemisphere (or hemicylinder) can be very different from the parameters of a jet formed by a whole sphere [15] or cylinder [16].

In the present paper we show that parameter of whispering gallery waves can be also quite different from the whole sphere or cylinder. Optimization of the remote segment thickness permits to create highly localized film distribution. This effect in cylinders can be used for contacting optical lithography with super resolution in the line thickness.

**Photonic nanojet: from geometrical optics approximation to the Mie theory**
A spherical transparent particle can focus light. This effect has been known for a few millennia: for example, Pliny the Elder (AD 23–AD 79) reported on the incendiary action of glass spheres [9]. This effect easy to understand under geometric optics approximation [17, 18]. The simplest way is to use ray tracing [19] and Snell's law [20, 21] (this technique has been known since Kepler). Refracted rays form a caustic, which is presented by parametric equation (here all coordinates $x$ and $y$ are normalized by particle radius, $R$):

$$x_c = \left[1 - \frac{1}{2}\frac{\sqrt{n^2-1+\cos^2\varphi}-2\cos\varphi}{\sqrt{n^2-1+\cos^2\varphi}-\cos\varphi}\cos\varphi\right]\cos\psi, \quad y_c = \sec\psi\sin\varphi + x_c\tan\psi, \tag{1}$$

where $\psi = 2\left[\varphi - \arcsin\left(\frac{\sin\varphi}{n}\right)\right]$.

Such caustic is presented by cuspoid curve [22,23], see in Fig. 1. The singularity point (geometrical optics focus [24]) situated at

$$x_s = x_c\big|_{\varphi\to 0} = \frac{n}{2(n-1)}. \tag{2}$$

Thus, $x_s \to \infty$ at $n \to 1$ and $x_s \to 1$ at $n \to 2$. When $n > 2$ the caustic is situated inside the sphere, this is the case of materials with a high refractive index. Such materials are used in optically resonant dielectric nanostructures [25], while materials with a refractive index of less than two [9] are the main materials for most optical components (lenses, optical fibers, etc.). This caustic was analyzed in a number of papers due to the problem of photonic nanojet, see Ref. [9] and references there. From the Eq. (1) one should define the numerical aperture $NA = n\sin\chi$, where the angle $\chi$ calculated at the surface of the particle, i.e. at $\varphi = \arccos\sqrt{\frac{n^2-1}{3}}$. This estimation yields $NA = n\sqrt{\frac{4-n^2}{3}}$. Maximal field enhancement in the focal point can be estimated as $I = \left[\sin\varphi_a/y_c(\varphi_a)\right]^2$, where $\varphi_a = \arccos\sqrt{(n^2-1)/3}$. Position of geometrical optics focus can be approximated by

$$x_f = \frac{2n^6 + 9n^4 + 48n^2 - 32}{6n^2(2+n^2)\sqrt{3(n^2-1)}}. \tag{3}$$

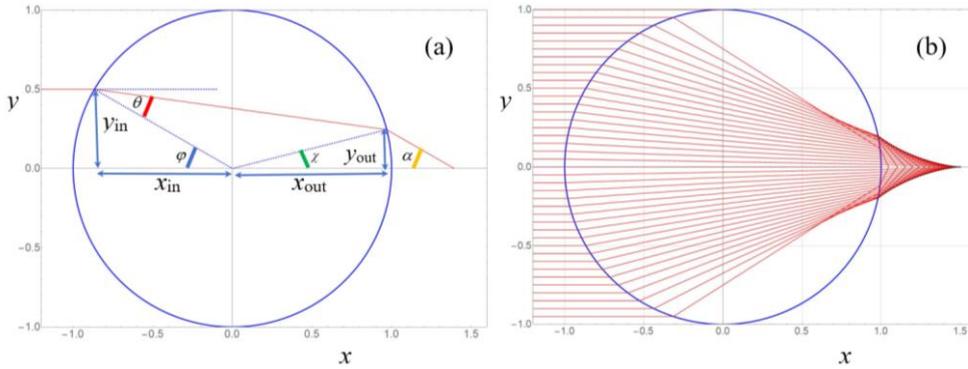

Fig. 1. (a) Ray tracing for a big particle with radius $R \square \lambda$. We introduce the incidence angle $\varphi$ and the refraction angle $\theta$ inside the sphere $\sin\varphi = n\sin\theta$. The ray enter into the particle at the point with coordinates $y_{in} = \tan\varphi$ and $x_{in} = -\cos\varphi$. The angles $\chi$ and $\alpha$ are given by $\chi = 2\theta - \varphi$ and $\alpha = 2\varphi - 2\theta$. Two close rays $y_c$ and $y_{cc}$ (corresponding to angles $\varphi$ and $\varphi + \delta\varphi$) emerged from the sphere after the second refraction are crossing at the caustic point $x_c = x_{out} + \partial_\varphi \sin\chi/\partial_\varphi \tan\alpha$. This yields the Eq. (1) for caustic. (b) The shape of the caustic from the Eq. (1) for the sphere with $n = 1.5$ is shown by dashed black line.

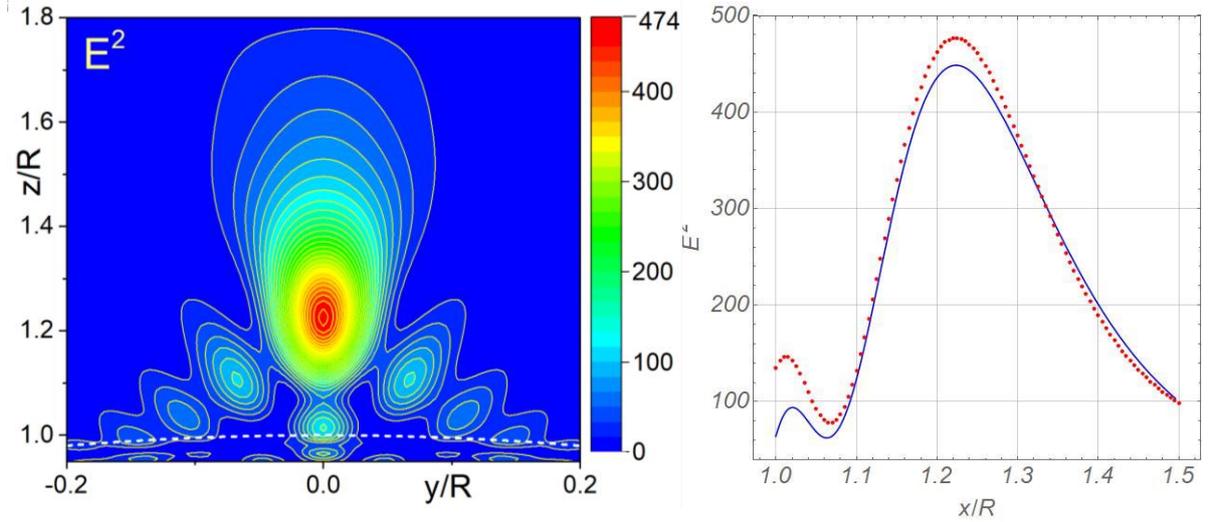

Fig. 2. (a) Distribution of intensity calculated from the Mie theory with $n=1.5$ and $q=70$. Such distribution is typical for Bessoid matching solution, see e.g. Fig. 5 in [24]. (b) Intensity distribution according to Bessoid approximation [24] (solid blue line) and from the Mie theory (dotted red line).

The solution with ray tracing yields just a qualitative picture of light focusing. One can compare this solution with the exact solution which follows from the Mie theory [6]. However the approximation of geometrical optics is valid just for sufficiently big size parameter $q = 2\pi R/\lambda \gg 1$ (here $R$ is radius of the particle and $\lambda$ is radiation wavelength). Thus, exact solution requires the summation of a large number of terms, $\ell_{max} = q + 4.05\, q^{1/3} + 2$ [26], in a multipole expansion even for moderate sphere sizes. For such big particles one can use the method of uniform caustic asymptotic [18]. The lowest nontrivial order correction for the field distribution is related to Bessoid integral. This integral appears naturally in the paraxial approximation [24]. Results of calculations are shown in Fig. 2.

According to the Mie theory [6,26] the scattering efficiency $Q_{sca} = \sigma_{sca}/(\pi R^2)$ (here $\sigma_{sca}$ is scattering cross section) can be expressed as a sum of electric $Q_\ell^{(e)}$ and magnetic $Q_\ell^{(m)}$ scattering efficiencies associated with the multipolar moment of $\ell$ th order multipoles:

$$Q_{sca} = \sum_{\ell=1}^{\infty}\left(Q_\ell^{(e)} + Q_\ell^{(m)}\right),\quad Q_\ell^{(e)} = \frac{2}{q_m^2}(2\ell+1)|a_\ell|^2,\quad Q_\ell^{(m)} = \frac{2}{q_m^2}(2\ell+1)|b_\ell|^2, \qquad (4)$$

where the electric $a_\ell$ and magnetic $b_\ell$ scattering amplitudes are defined by formulas

$$a_\ell = \frac{\mathfrak{R}_\ell^{(a)}}{\mathfrak{R}_\ell^{(a)} + i\mathfrak{I}_\ell^{(a)}},\quad b_\ell = \frac{\mathfrak{R}_\ell^{(b)}}{\mathfrak{R}_\ell^{(b)} + i\mathfrak{I}_\ell^{(b)}}, \qquad (5)$$

with $\mathfrak{R}_\ell^{(a,b)}$ and $\mathfrak{I}_\ell^{(a,b)}$ coefficients expressed by

$$\begin{aligned}
\mathfrak{R}_\ell^{(a)} &= n_p\psi_\ell(q_p)\psi_\ell'(q_m) - n_m\psi_\ell(q_m)\psi_\ell'(q_p),\\
\mathfrak{I}_\ell^{(a)} &= n_p\psi_\ell(q_p)\chi_\ell'(q_m) - n_m\chi_\ell(q_m)\psi_\ell'(q_p),\\
\mathfrak{R}_\ell^{(b)} &= n_p\psi_\ell(q_m)\psi_\ell'(q_p) - n_m\psi_\ell(q_p)\psi_\ell'(q_m),\\
\mathfrak{I}_\ell^{(b)} &= n_p\chi_\ell(q_m)\psi_\ell'(q_p) - n_m\chi_\ell'(q_m)\psi_\ell(q_p).
\end{aligned} \qquad (6)$$

Here the functions $\psi_\ell(z) = \sqrt{\frac{\pi z}{2}} J_{\ell+\frac{1}{2}}(z)$ and $\chi_\ell(z) = \sqrt{\frac{\pi z}{2}} N_{\ell+\frac{1}{2}}(z)$ are expressed through the Bessel and Neumann functions [11]. We use the subscripts $m$ and $p$ to denote the values referring to the external media and the particle, with refractive indices $n_m$ and $n_p$, respectively. In the expressions above, $q_m = q n_m$ and $q_p = q n_p$. The symbol $q$ represents the so-called size parameter, defined as $q = \omega R/c = 2\pi R/\lambda$.

The electric and magnetic fields inside the particle are expressed through the internal scattering amplitudes $c_\ell$ and $d_\ell$ given by [6]

$$c_\ell = \frac{i n_p}{\Re_\ell^{(a)} + i \Im_\ell^{(a)}}, \quad d_\ell = \frac{i n_p}{\Re_\ell^{(b)} + i \Im_\ell^{(b)}}. \quad (7)$$

Although the denominators of these amplitudes are the same as in amplitudes $a_\ell$ and $b_\ell$ in (5), which means that position of these resonances are close, the numerators of (7) never tends to zero. As a result the values of amplitudes $|c_\ell|^2$ and $|d_\ell|^2$ are not restricted by unity as amplitudes $|a_\ell|^2$ and $|b_\ell|^2$ in (2), but increase with values of size parameter and refractive index. To compare amplitudes it is convenient to introduce partial internal scattering efficiencies, similar to those in Eq. (1):

$$F_\ell^{(e)} = \frac{2}{q_m^2}(2\ell+1)|c_\ell|^2, \quad F_\ell^{(m)} = \frac{2}{q_m^2}(2\ell+1)|d_\ell|^2. \quad (8)$$

It leads to specific variation of $c_\ell$ and $d_\ell$ amplitudes at the big $\ell$ values. Namely, the amplitudes of these functions are quite small up to some threshold values, $q < q_{tr}$, which are of the order of $\ell$ [27], see an example in Fig. 3. The first narrow resonance at $q = q_{tr}$, where $q_{tr} \cong \ell$, plays a dominant role in the Mie theory. The total electric field $\mathbf{E}$ (similar in $\mathbf{H}$) can be presented as a sum of a single resonant term $\mathbf{E}_{\ell=\ell_{res}}$ with a narrow spectrum and the field from all other nonresonant terms $\sum_{\ell \neq \ell_{res}} \mathbf{E}_\ell$ with a broad spectrum. The interference of the signals with a broad and narrow spectums yields the Fano resonance, which produces the narrow resonances in the intensity. These resonances in electric and magnetic fields greatly exceed corresponding resonances in the scattering efficiency [10].

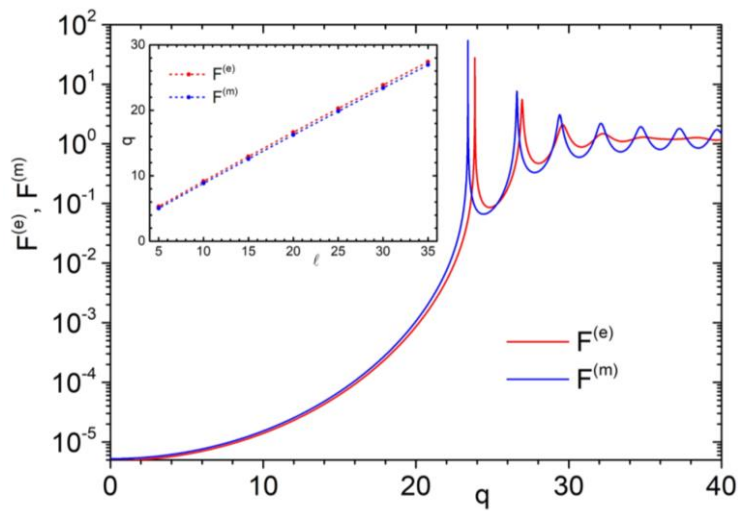

Fig. 3. Amplitudes $|F_\ell^{(e)}|$ and $|F_\ell^{(e)}|$ for $\ell = 30$ and $n_p = 1.5$ versus size parameter $q$. The first sharp resonance arise at $q$ by the order of $\ell$. Insert shows how the position of the first sharp resonance vary with $\ell$ number.

The electric and magnetic fields can be defined through the derivations of the electric, $\Pi^{(e)} = \sum \Pi_\ell^{(e)}$, and magnetic, $\Pi^{(m)} = \sum \Pi_\ell^{(m)}$, Debye potentials. Inside the particle these potentials are presented by [17]

$$\Pi_\ell^{(e)} = c_\ell \frac{i^{\ell+1}}{k^2} \frac{2\ell+1}{\ell(\ell+1)} \psi_\ell(kr) P_\ell^{(1)}(\cos\theta)\cos\varphi,$$

$$\Pi_\ell^{(m)} = d_\ell \frac{i^{\ell+1}}{k^2} \frac{2\ell+1}{\ell(\ell+1)} \psi_\ell(kr) P_\ell^{(1)}(\cos\theta)\sin\varphi. \qquad (9)$$

Here $k$ is the wave vector inside the particle, $kR = q_p$.

From the formulas (9) one explains the physical origin of the whispery gallery waves (WGW) formation [28]. Argument spherical Bessel function $\psi_\ell(kr)$ varies from zero in the center of the particle till parameter $q_p$ on the surface of the particle. Thus, at big $\ell$ values $\psi_\ell$ function is close to zero till to critical value $q_p$ which is by the order of $\ell$, see in Fig. 4a. If the first zero of the Bessel function is located near the radius of the sphere, then almost the entire field of this wave will be located in a very narrow region near the surface of the sphere. This is the WGW case. The angular field modulation is presented by Legendre function, see in Fig. 4b.

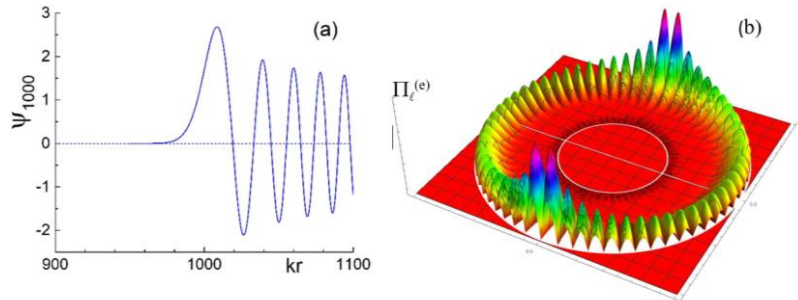

Fig. 4. (a) Spherical Bessel function $\psi_\ell(kr)$ at big index $\ell = 1000$ [28]. (b) Spacial distribution of the modulus of the $\Pi_\ell^{(e)}$ function (9) at $\varphi = 0$ and $\ell = 30$ for $n = 1.5$ and $q = 23.855$.

The WGW can be interpreted as a wave of total internal reflection of the ray propagating along the surface of the dielectric sphere [28]. However, if the interface has a curvature (as in the case of the sphere), then the internal reflection is not complete: part of the wave still seeps out of the ball out. One can see this leakage in Fig. 4b. It occurs through the white triangles situated on the surface of the particle. The leakage is the smaller, the larger the radius of the ball in comparison with the wavelength. That is, for the existence of WGW, the sphere must have a rather sick radius in comparison with the wavelength of light, i.e. it needs the condition $q_p \Box 1$. Calculations with the Mie theory show that the WGW structure in the electric field intensity can be seen even at the size parameter by the order of ten, see in Fig. 5.

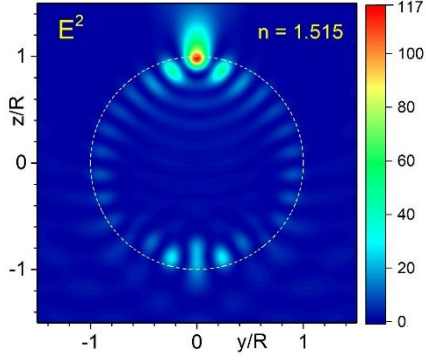

Fig. 5. Distribution of electric intensity $E^2$ within the $yz$ plane of the particle with refractive index $n = 1.515$ and size parameter $q = 11$.

**Janus particle: from geometrical optics to the whispery gallery waves**

One can examine the light focusing by truncated cylinder or a sphere, see in Fig. 6. Truncated segment can be replaced by a similar portion of different material – such structures are called as Janus particles [29]. We use notation $h$ for the height of truncated segment. In particular case replaced portion can be an air or vacuum. The ray tracing technique applied for truncated sphere yields the following shape of the caustics:

$$z_c = 1 - h - \frac{\sec \beta}{n} \frac{(1 - n^2 \sin^2 \beta)^{3/2}}{\cos \varphi - \sqrt{n^2 - \sin^2 \varphi}} \tag{10}$$

$$\left( \cos\varphi \sqrt{n^2 - \sin^2\varphi} + (1 - h + \cos\varphi)\sec^2\beta \left(\cos\varphi - \sqrt{n^2 - \sin^2\varphi}\right) + \sin\varphi\sqrt{n^2 - \sin^2\varphi}\tan\beta \right),$$

$$y_c = y_{out} - (z_c - 1 + h)\tan\gamma, \tag{11}$$

$$y_{out} = \sin\varphi - (1 - h + \cos\varphi)\tan(\varphi - \theta), \quad \gamma = \arcsin[n\sin(\varphi - \theta)], \quad \beta = \varphi - \theta, \quad \theta = \arcsin\left(\frac{\sin\varphi}{n}\right).$$

The singularity point this Janus particle situated at

$$z_f = z_c \big|_{\varphi \to 0} = 1 - \frac{2}{n} + \frac{1}{n-1} + h\left(\frac{1}{n} - 1\right). \tag{12}$$

The basic effect with Janus particle is related to variation of focal length in comparison with the initial sphere. For example, the sphere with $n = 1.33$ has the focal point situated at $z_f = 2.015$. After small truncation with $h = 0.07$ corresponding Janus particle has a focal point with $z_f = 2.509$, i.e. truncation yields the longer focus. The ray tracing technique [19] approximation gave qualitatively correct description of focusing properties of transparent dielectric.

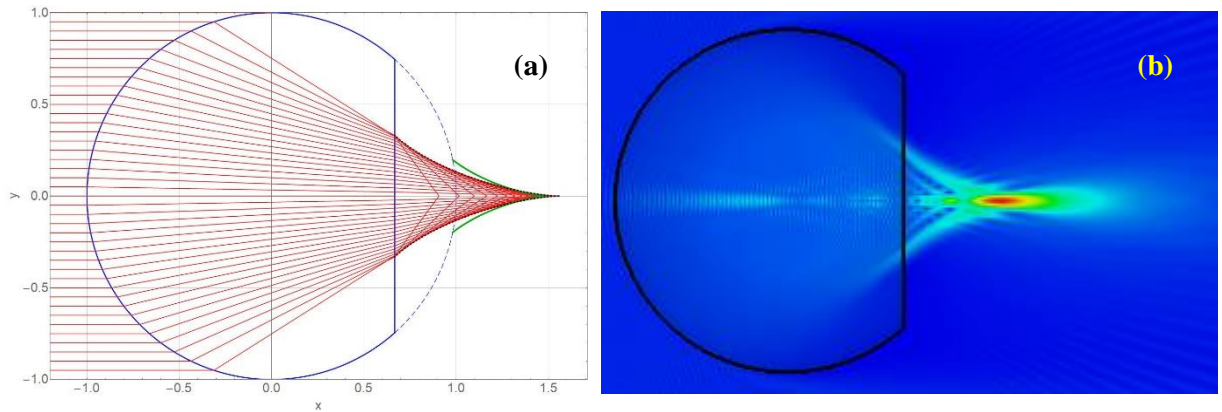

Fig. 6. (a) We introduce the same incidence angle $\varphi$ and the refraction angle $\theta$ as in Fig. 1a. Here $h$ is the height of truncation normalized to particle radius $R$. Ray emerges from the sphere after the second refraction with the angle $\gamma$ follows the Snell's law $\sin\gamma = n\sin(\varphi-\theta)$. The shape of the caustic for the truncated sphere with $h = 1-1/n$ and $n = 1.5$ is shown by dashed black line. The solid green line shows the caustic of the spherical particle with the same refractive index. (b) The same parameters and the exact solution of the Maxwell equation, corresponding to size parameter $q = 2\pi R/\lambda = 100$.

The truncated particle presents a solid immersion lens (SIL) which can overcome diffraction limit [13]. It could be noted that physical principles of truncated spherical SIL, for which aberration free focusing occurs and also known as Weierstrass SIL, is based on compressing the emitted light into a small NA by decreasing of the refraction angle of the transmitted light, measured from the optical axis. This occurs when the sphere is truncated to a thickness h=r(l+ l/n), where r is the radius of the sphere, and h corresponds to the aplanatic focus, see in [17] (p.253) and [30]. Similar SIL have been used in optical microscopes and photolithography [31]. Analysis of the photonic nanojet with truncated spherical particle shows that the maximal intensity in the focal point is less than those produced by spherical particle, but the effective focal length can be much greater [15].

We repeated a similar numerical analysis for a different design of Janus particles and found some new effects. As an example, we show in Fig. 7 the cylindrical Janus particle consisting from two half cylinders with different refractive index: $n = 1.5$ for the bottom part and $n = 1.3$ for the upper part. The radius of a whole cylinder is equal to $R = 5\lambda$. Thus, the size parameter $q = 5\pi \square 1$. At normal condition one can see a usual photon nanojet with maximal field enhancement $E^2 = 17$ at the focal point. A small truncation with $h = d/R = 0.03$ leads to the field redistribution due to strong WGW excitation with maximal field enhancement $E^2 = 23.5$. A further truncation with $h = 0.04$ practically restore the initial photon nanojet. A similar behavior can be seen for magnetic intensity variation.

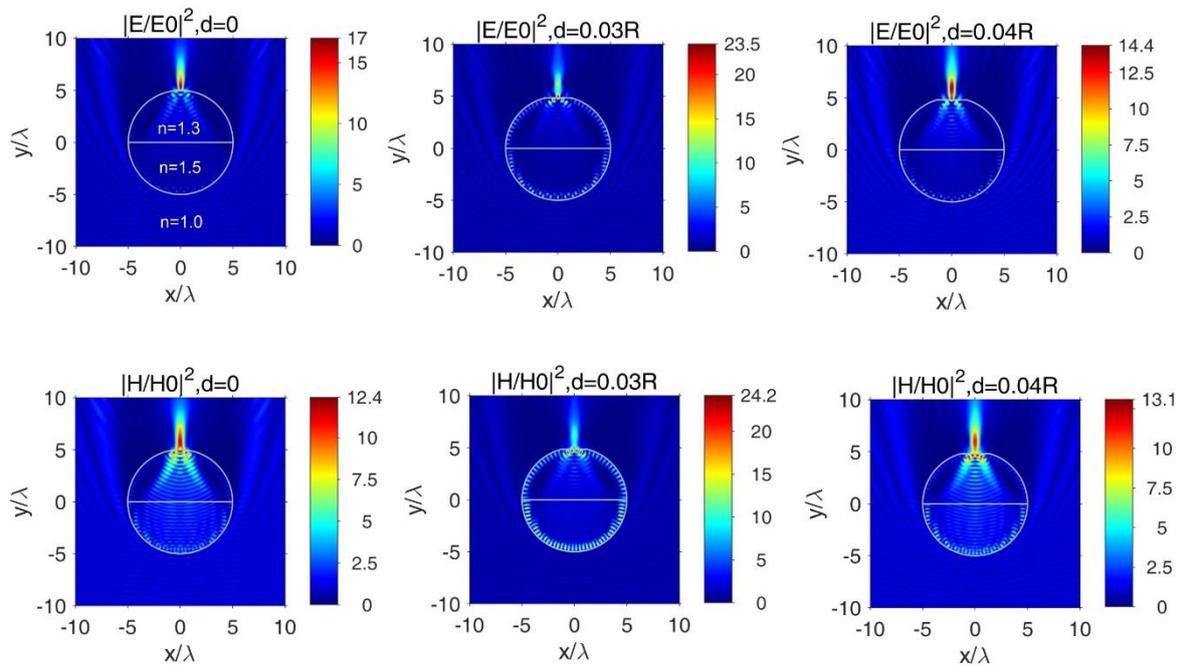

Fig. 7. Distribution of electric $E^2$ intensity (picures on the top) and magnetic $H^2$ intensity (down pictures) within the cross section of the Janus cylinder with refractive index $n = 1.5$ (down), $n = 1.3$ (top), and size parameter $q = 5\pi$.

An example of truncation effect for the intensity distribution is shown in Fig. 7. Namely, for the cylinder with refractive index $n=1.5$ and big size parameter $q=2\pi R/\lambda =100$ we can see a number of oscillations with maximal intensity around the particle. These oscillations depend on light polarization and the depth of truncated layer. Amplitudes of these resonances are typically about 10% of average intensity with field enhancement about 25. However, for some small truncation with $d=0.015R$ we can see in Fig. 8 resonance for TM polarized light with field enhancement in the hot spots on the surface. It looks like resonant excitation of surface electromagnetic wave within the plane disk of the truncated surface. The singularity related to phase discontinuity at the line where the spherical (or cylindrical) surface cross the plane surface leads to change of the Snell's law to generalized laws of reflection and refraction [32]. According to this law occurs the variation of critical angles for total internal reflection. At some value of phase gradient there is a critical angle of incidence above which the reflected beam becomes evanescent [32]. Under the approximation of geometrical optics this phase gradient depends on the thickness $h$ of truncated element and refractive index $n$. In Fig. 9c one can see the result of interference of two evanescent waves. We also found that the efficiency of the excitation of whispery gallery waves strongly depends on the $h$ value.

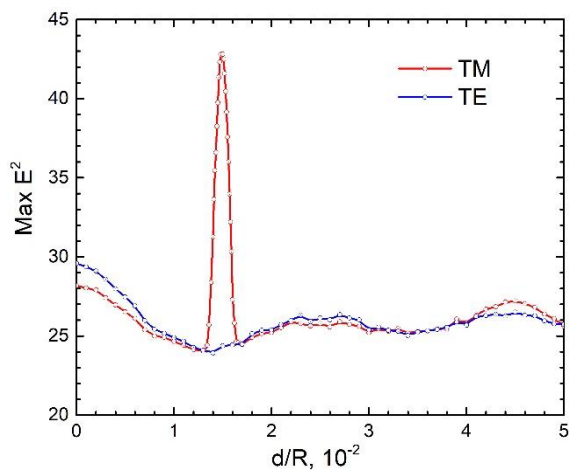

Fig. 8. Maximal field enhancement around the truncated cylindrical versus the depth of truncated element.

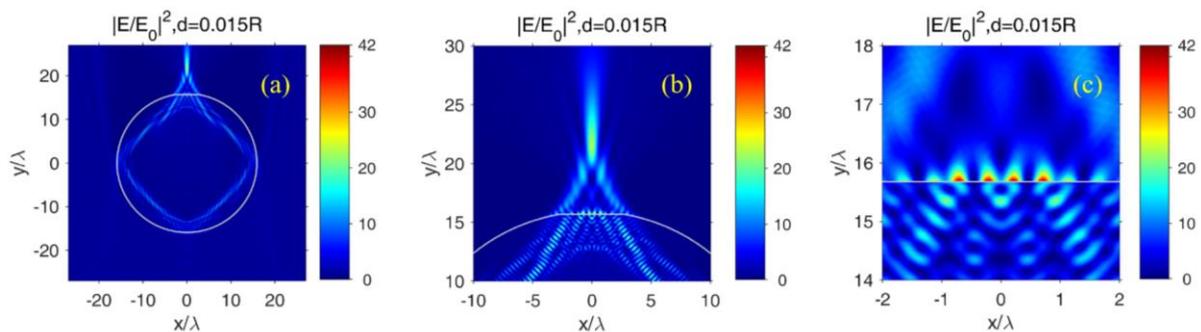

Fig. 9. Distribution of the field intensity for a resonant value of truncation (a) and zoom in (b) and (c).

Optimization of this effect for cylinders permits to reach a super resolution in the line thickness what can be used for contacting optical lithography

**Truncated nanowires for near field optical lithography**
The most powerful modern lithographic technique is related to usage of focused electron beams [33] or ion beams [34] (e.g. a focused helium ion beam [35]). These techniques permit to realize patterning fidelity at nanometer scale dimensions. However, beam technology is associated with very expensive and complex equipment. In addition, it is rather slow.

Over the past decades, a number of new ideas have been proposed to create a fast-lithographic technique that allows mass production of structures with a scale of tens of nanometers. Among these ideas, for example, an evanescent interferometric lithography [36-38] can be mention. Other techniques involve laser induced tip-assisted Atomic Force Microscope (AFM) [39] or a Near-field Scanning Optical Microscope (NSOM) [40,41]. Finally, we should mention 'plasmon printing' technology [42-44]. The later relies on the surface plasmon resonance occurring in metal nanoparticles, which can produce sub-wavelength structures.

Returning to lithography using Janus particles, we note once again the resonance properties of small truncation od the cylinder, see in Fig. 10. Here we show the maximal intensities inside and outside particle. Each resonance corresponds to resonant excitation of whispery gallery waves. Similar resonances existing within the nontruncated particles as well (see e.g. movie in supplementary materials in Ref. [9]). The difference is that in truncated particles these resonances are sharper. We also draw attention to the fact that in Janus particles the magnetic nanojet mode appears to be more contrasting compare to nontruncated particles [9]. It gave the additional functionality to work with magnetic materials.

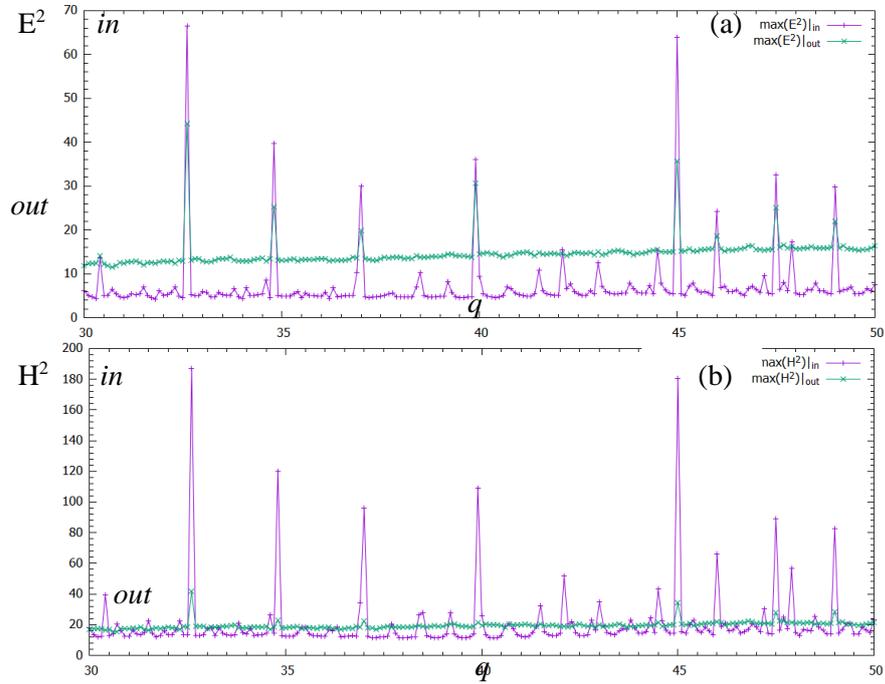

Fig. 10. Internal and external electric (a) and magnetic (b) intensities versus size parameter for the cylinder with fixed truncation parameter $h = 0.02$. Size parameter $q = 32.5$ correspnds to radius of the cylinder $R \approx 10\lambda$.

Depending on truncation parameter $h$ and radiation wavelength one can realize situation with a single, double ore four-maximums distributions in the intensity on the plane truncated surface. Having the spacing between cylinders as additional parameter we can imagine lithographic technique schematically presented in Fig.11. The number of cylinders with refracted index $n_p$ are embedded into matrix with refracted index $n_m < n_p$. It is equivalent to situation when the cylinders with relative index $n = n_p/n_m$ are cituated in vacuum. The surface of this matrix can be polished to reach a presize truncation.

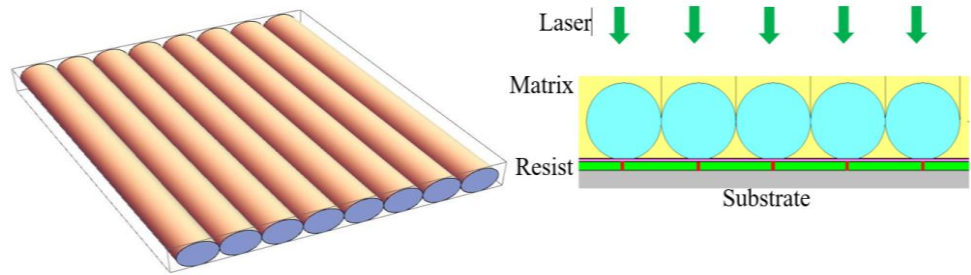

Fig. 11. Schematic for a lithographic process with truncated cylinders. Here an important role plays a thin protected layer between the matrix and photoresist, which plays a role of anti-reflective coating, depending on the thickness of the coating.

We performed a modeling of this technique to analyze the patterning fidelity at nanometer scale dimensions. The length of the lines can be on a centimeters' scale.

**References**

[1] T. Ito, S. Okazaki, *Pushing the limits of lithography*, Nature **406**, 1027-1031 (2000).
[2] Y. F. Lu, L. Zhang, W. D. Song, Y. W. Zheng, B. S. Luk'yanchuk, *Laser Writing of a Subwavelength Structure on Silicon (100) Surfaces with Particle-Enhanced Optical Irradiation*, JETP Lett. **72**, 457–459 (2000).
[3] M. Mosbacher, H.-J. Münzer, J. Zimmermann, J. Solis, J. Boneberg, P. Leiderer, *Optical field enhancement effects in laser-assisted particle removal*, Appl. Phys. A 72, 41–44 (2001).
[4] H.-J. Münzer, M. Mosbacher, M. Bertsch, J. Zimmermann, P. Leiderer, J. Boneberg, *Local field enhancement effects for nanostructuring of surfaces*, J. Microscopy **202**, 129-135 (2001).
[5] K. Piglmayer, R. Denk, D. Bäuerle, *Laser-induced surface patterning by means of microspheres*, Appl. Phys. Lett. **80**, 4693-4695 (2002).
[6] C.F. Bohren and D.R. Huffman, *Absorption and Scattering of Light by Small Particles*. Wiley, 1998.
[7] P. A. Bobbert, J. Vlieger, *Light scattering by a sphere on a substrate*, Physica A **137**, 209-242 (1986).
[8] B. S. Luk'yanchuk, Y. W. Zheng, Y. F. Lu, *Laser cleaning of solid surface: optical resonance and near-field effects*, Proc. SPIE **4065**, 576-587 (2000).
[9] B.S. Luk'yanchuk, R. Paniagua-Domínguez, I. V. Minin, O. V. Minin, Z.B. Wang, *Refractive index less than two: photonic nanojets yesterday, today and tomorrow*. Opt. Mat. Express **7**, 1820-1847 (2017).
[10] Z.B. Wang, B. Luk'yanchuk, L. Yue, B. Yan, J. Monks, R. Dhama, O.V. Minin, I.V. Minin, S.M. Huang, A.A. Fedyanin, *High order Fano resonances and giant magnetic fields in dielectric microspheres*. Scientific Reports **9**, 20293 (2019).
[11] A. Walther, A.H.E. Müller, *Janus particles*, Soft Matter **4**, 663-668 (2008).
[12] J. Hu, S. Zhou, Y. Sun, X. Fanga, L. Wu, *Fabrication, properties and applications of Janus particles*, Chem. Soc. Rev. **41**, 4356–4378 (2012).
[13] D.R. Mason, M.V. Jouravlev, K.S. Kim, *Enhanced resolution beyond the Abbe diffraction limit with wavelength-scale solid immersion lenses*, Optics Letters **35**, 2007-2009 (2010).
[14] G.J. Lee, H.M. Kim, Y.M. Song, *Design and Fabrication of Microscale, Thin-Film Silicon Solid Immersion Lenses for Mid-Infrared Application*, Micromachines **11**, 250 (2020).
[15] C.-Y. Liu, *Photonic nanojet shaping of dielectric non-spherical microparticles*, Physica E **64**, 23–28 (2014).

[16] G. Gu, X. Shen, Z. Peng, X. Yang, S. Bandyopadhyay, J. Feng, D. Xiao, L. Shao, *Numerical investigation of photonic nanojets generated from D-shaped dielectric microfibers*, Proc. SPIE **11186**, 111861H (2019).

[17] M. Born, E. Wolf, *Principles of optics: electromagnetic theory of propagation, interference and diffraction of light*. Elsevier, 2013.

[18] Y.A. Kravtsov, Y.I. Orlov, *Caustics, catastrophes and wave fields*. Springer Science & Business Media, 2012.

[19] A.S. Glassner, *An Introduction to Ray Tracing*. Academic Press, 1991.

[20] B.S. Luk`yanchuk, N. Arnold, S.M. Huang, Z.B. Wang, M.H. Hong, *Three-dimensional effects in dry laser cleaning.* Appl. Phys. A **77**, 209-215 (2003).

[21] N. Arnold, *Theoretical description of dry laser cleaning.* Appl. Surf. Sci. **208**, pp. 15–22 (2003).

[22] V.I. Arnol'd, *Catastrophe theory*. Springer Science & Business Media, 2003.

[23] T. Poston, I. Stewart, *Catastrophe theory and its applications*. Courier Corporation, 2014.

[24] J. Kofler, N. Arnold, *Axially symmetric focusing as a cuspoid diffraction catastrophe: Scalar and vector cases and comparison with the theory of Mie*. Phys. Rev. B **73**, 235401 (2006).

[25] A.I. Kuznetsov, A.E. Miroshnichenko, M.L. Brongersma, Y.S. Kivshar, B. Luk`yanchuk, *Optically resonant dielectric nanostructures*. Science, vol. **354**, aag2472 (2016).

[26] P.W. Barber, S.C. Hill, *Light Scattering by Particles: Computational Methods*. Word Scientific, 1990.

[27] Z.B. Wang, B. Luk'yanchuk, L. Yue, R. Paniagua-Dominguez, B. Yan, J. Monks, R. Dhama, O.V. Minin, I.V. Minin, S.M. Huang, A.A. Fedyanin, *Super-resonances in microspheres: extreme effects in field localization,* arXiv:1906.09636 (2019).

[28] A.N. Oraevsky, *Whispering-gallery waves*. Quantum Electronics **32**, 377 (2002).

[29] P. Liu, A. T. Liu, D. Kozawa, J. Dong, J. F. Yang, V. B. Koman, M. Saccone, S. Wang, Y. Son, M. H. Wong, M. S. Strano, *Autoperforation of 2D materials for generating two-terminal memristive Janus particles*, Nature Materials **17**, pp. 1005-1012 (2018).

[30] G. S. Kino, S. M. Mansfield, *Near field and solid immersion optical microscope*. US Patent 5,004,307 (1991).

[31] L. P. Ghislain, V. B. Elings, K. B. Crozier, S. R. Manalis, S. C. Minne, K. Wilder, G. S. Kino, C. F. Quate. *Near-field photolithography with a solid immersion lens*. Appl. Phys. Lett. **74**, 501 (1999).

[32] N. Yu, P. Genevet, M.A. Kats, F. Aieta, J.-P. Tetienne, F. Capasso, Z. Gaburro. *Light Propagation with Phase Discontinuities: Generalized Laws of Reflection and Refraction*. Science **334**, pp. 333-337 (2011).

[33] Y. Chen, *Nanofabrication by electron beam lithography and its applications: A review*, Microelectronic Engineering **135**, 57-72 (2015).

[34] F. Watt, A. A. Bettiol, J. A. van Kan, E. J. Teo, M.B.H. Breese, *Ion beam lithography and nanofabrication: A review*, Int. J. Nanoscience **4**, 269–286 (2005).

[35] A.I. Kuznetsov, A.E. Miroshnichenko, Y.H. Fu, V. Viswanathan, M. Rahmani, V. Valuckas, Z.Y. Pan, Y. Kivshar, D.S. Pickard, B. Luk`yanchuk, *Split-ball resonator as a three-dimensional analogue of planar split-rings*, Nature Communications **5**, 3104 (2014).

[36] C.S. Lim, M.H. Hong, Y. Lin, Q. Xie, B.S. Luk'yanchuk, A.S. Kumar, M. Rahman, *Microlens array fabrication by laser interference lithography for super-resolution surface nanopatterning*, Appl. Phys. Lett. **89**, 191125 (2006).

[37] Y. Zhou, M.H. Hong, J. Y. H. Fuh, L. Lu, B.S. Lukiyanchuk, *Evanescent wave interference lithography for surface nano-structuring*, Physica Scripta **T129,** 35-37 (2007).

[38] K.V. Sreekanth, J.K. Chua, V.M. Murukeshan, *Interferometric lithography for nanoscale feature patterning: a comparative analysis between laser interference, evanescent wave interference, and surface plasmon interference*, Applied Optics **49**, 6710-6717 (2010).

[39] Z.B. Wang, J. Naveen, L. Li, B.S. Luk'yanchuk, *A Review of Optical Near-Fields in Particle/Tip-assisted Laser Nanofabrication*, Journal of Mechanical Engineering Science **224**, 1113-1127 (2010).

[40] W. Wang, M. H. Hong, D. Wu, Y. W. Goh, Y. Lin, P. Luo, B. Luk'yanchuk, T. C. Chong, *Ultrafast laser nanofabrication assisted with near field scanning optical microscopy*, Proc. SPIE **5063**, 449-453 (2003).

[41] A.A. Tseng, *Recent developments in nanofabrication using scanning near-field optical microscope lithography*, Optics & Laser Technology **39**, 514–526 (2007)

[42] P.G. Kik, A.L. Martin, S.A. Maier, H.A. Atwater, *Metal nanoparticle arrays for near field optical lithography*, Proc. SPIE **4810**, 7-13 (2002).

[43] Z. B. Wang, M. H. Hong, B. S. Luk`yanchuk, S. M. Huang, O. F. Wang, L. P. Shi, T. C. Chong, *Parallel nanostructuring of GeSbTe films with particle-mask* Applied Physics A **79**, 1603 –1606 (2004).

[44] D. Eversole, B. Luk'yanchuk, A. Ben-Yakar, *Plasmonic laser nanoablation of silicon by the scattering of femtosecond pulses near gold nanospheres*, Appl. Phys. A **89**, 283–291 (2007).